\documentclass[onecolumn]{aa}
\usepackage{graphicx}
\usepackage{indentfirst}
\usepackage{natbib}

\begin{document}
   \title{The VIRMOS deep imaging survey: I. overview and survey strategy}


   \author{O. Le F\`evre \inst{1}, Y. Mellier \inst{2,3}, 
          H.J. McCracken \inst{1,4}, S. Foucaud \inst{1}, S. Gwyn \inst{1},
         M. Radovich \inst{2,5}, M. Dantel-Fort \inst{3}, E. Bertin \inst{2,3}
         C. Moreau \inst{1}, J.-C. Cuillandre
         \inst{6}, M. Pierre \inst{7},
         V. Le Brun \inst{1}, A. Mazure \inst{1},  L. Tresse \inst{1} 
         \fnmsep\thanks{The data presented in this paper has been 
         obtained with the Canada-France-Hawaii Telescope, operated by the 
         National Research Council of Canada, the Centre National de la
         Recherche Scientifique of France, and the University of Hawaii }
          }

  \authorrunning{O. Le F\`evre, Y. Mellier, H.J. McCracken, et al.}
  \titlerunning{VIRMOS imaging survey: I. Overview and strategy}

   \offprints{O. Le F\`evre}

   \institute{Laboratoire d'Astrophysique de Marseille, UMR 6110, Traverse 
    du Siphon-Les trois Lucs, 13012 Marseille, France\\
              email: olivier.lefevre@oamp.fr
         \and
Institut d'Astrophysique de Paris, UMR 7095, 98 bis Bvd Arago, 75014 Paris, France
\and
Observatoire de Paris, LERMA, UMR 8112, 61 Av. de l'Observatoire, 75014 Paris, France
\and
Osservatorio Astronomico di Bologna, via Ranzani 1, 40127 Bologna, Italy
\and
Osservatorio di Capodimonte, via Moiariello 16, 80131 Napoli, Italy
\and
Canada-France Telescope Corporation, 65-1238 Mamalahoa Hwy, Kamuela, Hawaii 96743, USA
\and
Service d'Astrophysique, CE Saclay, L'Orme des Meurisiers, 91191 Gif sur Yevette Cedex , France
             }

   \date{Received May XX, 2003; accepted ..., 2003}

   \abstract{
  This paper presents the CFH12K-VIRMOS survey: 
a deep B, V, R and I imaging survey in four fields 
totalling more than 17 deg$^2$, conducted with the
$30\times40$ arcmin$^2$ field CFH-12K camera.
The survey is intended to be a multi-purpose survey used for a variety of
science goals, including surveys of very high redshift galaxies
and weak lensing studies. 

Four high galactic latitude fields, each  $2\times2$ deg$^2$, have been 
selected along the celestial equator:
0226-04, 1003+01, 1400+05, and 2217+00.  
The 16 deg$^2$ of the "wide" survey are 
covered with exposure times of 2h, 1.5h, 1h, 1h , 
while the $1.3\times1$ deg$^2$ 
area of the "deep" survey at the center of the 0226-04 field is covered with 
exposure times of 7h, 4.5h, 3h, 3h, in B,V,R and I respectively. 
An additional area $\sim2$deg$^2$ has been imaged in the
0226-04 field corresponding to the area surveyed by the 
XMM-LSS program \citep{pierre03}.

The data is pipeline processed at the Terapix facility at the Institut 
d'Astrophysique de Paris to produce large mosaic images. 
The catalogs produced contain the positions, shape, total and
aperture magnitudes for the 2.175  million objects  measured in the 4 areas
The depth measured as a $3\sigma$ measurement in a 3 arc-second aperture
is $I_{AB}=24.8$ in the ``Wide'' areas, and $I_{AB}=25.3$
in the deep area. Careful quality control has been applied 
on the data to ensure internal consistency and assess the
photometric and astrometric accuracy as described in 
joint papers \citep{mccracken03}.

These catalogs are used to select targets for the VIRMOS-VLT Deep Survey,
a large spectroscopic survey of the distant universe (Le F\`evre et al., 2003).
First results from the CFH12K-VIRMOS survey have been published on
weak lensing (e.g. van Waerbeke \& Mellier 2003).

Catalogs and images are available through the VIRMOS 
database environment under Oracle ({\tt http://www.oamp.fr/virmos}). 
They will be open for
general use on July 1st, 2003.

   }

   \maketitle
%

\section{Introduction}

Deep imaging over large areas is required for many
fields of astronomy, to survey large numbers of objects 
or to search for rare, low projected density objects. 
It is a key tool to our understanding
of the universe, from solar system studies to the most distant galaxies.
The measurement of positions, magnitudes, colors, and shape
are key observable to all astronomical investigations.
From astrometric and photometric measurements, a census of 
the number and positions of various classes of objects 
can be conducted as a function of magnitude, color, or shape. 
 
Progress in detector area have allowed to design
and build large CCD mosaics covering a significant fraction
of the field available on wide field telescopes
 \cite{groom00}. These mosaics have enabled 
survey work on large areas which had only been possible
previously with photographic plates at the focus of
Schmidt and prime foci of major telescopes. Probing
scales larger than 0.5 degree in one single exposure
with the sensitivity of CCDs 
has become possible at a few facilities
in the past few years \citep{cuillandre00,boulade00, miyazaki02,kuijken02}, 
 enabling a wealth of new deep survey 
initiatives \citep{postman98,nonino99,mccracken01,mcmahon01,wilson03}

This paper is the first of a serie that describes the 
deep survey we have undertaken starting when the CFH12K camera became
available at CFHT prime focus in 1999. The main survey covers a total of 
 16 deg$^2$
in 4 areas, each $2\times2$deg$^2$, which have been imaged in B,V,R
and I bands, at a depth equivalent to $I_{AB}=24.8$ 
for a $3\sigma$ detection in a 3 arc-second circular 
aperture for all the area, with a deeper area imaged to
$I_{AB}=25.3$.
We focus here on the survey goals and strategy, 
and the observations performed, together with
an overview of the pipeline 
processing and the content of the 
photometric catalogs.
 Other papers in this series \citep{mccracken03}; (Gwyn et al., 2003 in 
  preparation) 
describe in details the procedures
followed to build the astrometry and photometry of the large image mosaics,
and the quality control applied.  The VIRMOS survey has been 
  used already for cosmic shear studies and 8 refereed papers have been 
    published so far from this data set by the {\sc virmos-descart} 
     project  \citep{vanwearmel03}.

\section{Survey Goals}

The survey has been designed to 
address a broad range of astrophysical questions in one single observing
strategy.  Several main science drivers have been identified:
\begin{itemize}
\item study the evolution of galaxies from
redshifts $\sim5$, 
\item study the evolution of large scale structures 
over 100 $h^{-1}$ Mpc
from redshifts $\sim5$, 
\item measure weak lensing signature of large scale structures,
\item measure the properties of galaxy biasing using together the 
 dark matter (from weak lensing) and the galaxy distribution,
\item identify new high redshift clusters of galaxies,
\item identify faint AGN and study their evolution, 
\item identify new Kuiper belt objects to
give new insights into the formation of the solar system, 
\item identify very faint halo white dwarfs, 
\item study high redshift Lyman-break galaxies with $3.5<z<4.2$. 
 using the multi-color data set of the optical and U-band survey 
\end{itemize}
The data is also being used with time series 
to search for high redshift supernovae.

This imaging survey is being used to select the targets 
of the VLT-VIRMOS deep redshift survey of more than 100000 galaxies with
$0<z<5+$ \citep{lefevre03}, to 
  obtain large catalogues of galaxy shapes for cosmic shear studies 
    and as the optical
counterpart to the XMM medium deep survey 
 ({\tt http://vela.astro.ulg.ac.be/themes/spatial/xmm/LSS/})
  being carried out
with XMM \cite{pierre03}. The 
 deep VIRMOS field is also completed by a U-band follow up
  done at ESO, with the WFI camera (Radovich et al 2003), as well as 
    in radio wavelength with the  VLA  \cite{bondi03}.  A 
      tiny area of the deep field has also been observed in 
       K-band, with SOFI at the ESO/NTT (Iovino et al 2003).

\section{Survey Strategy}

\subsection{Field Selection}

\begin{figure}
\resizebox{\hsize}{!}{\includegraphics{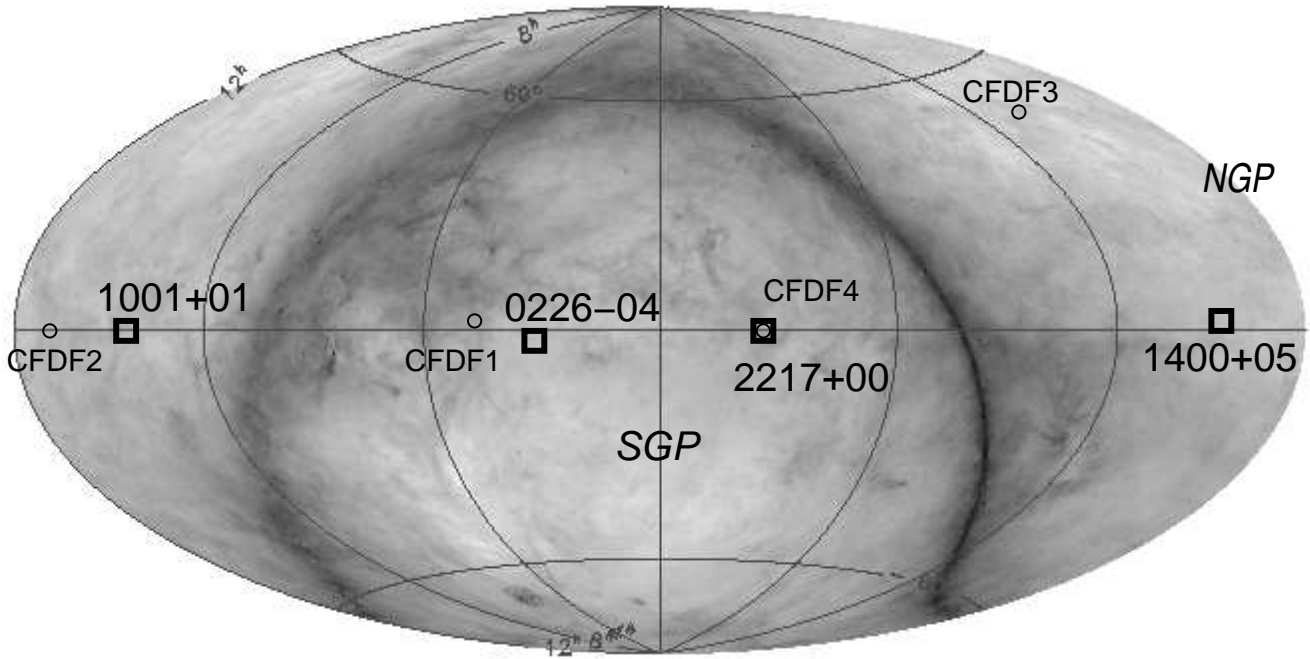}}
   \caption[]{Positions of the 4 VIRMOS fields on a dust map
    \citep{schlegel98} of the sky
    (Aitoff projection in equatorial coordinates). 
     The open square are the center positions. One can see that the 
      four fields are visible from CFHT as well as the ESO Paranal 
       and the ESO La Silla observatories and are separated by at least 
         4 hours, making the observations optimally spread over the 
	   nights and over the year.
     The square size are not exactly scaled according to the true 
        field size (2$^o$$\times$2$^o$) to ease visualization. For comparison 
 the open circles indicates of center position of the CFDF fields
  \citep{mccracken01}. The CFDF field size are 16 times smaller 
  than the VIRMOS. The dust map was generated using the 
   advanced 
     Skyview virtual observatory 
      tools ({\tt http://http://skys4.gsfc.nasa.gov/}).
      }
\hspace{5cm}
        \label{virmosindust}
\end{figure}

Survey fields have been selected with the following criteria: 
\begin{itemize}
\item along the celestial equator
to allow for visibility from northern and southern hemispheres
observatories, 
\item galactic latitude higher
than $l=45$ deg., 
\item low cirrus absorption as seen from the DIRBE maps available when the 
survey started in 1998, 
\item visibility of any 2 fields at any time of the year to fill observing
nights. 
\end{itemize}
The four fields selected are listed in Table \ref{fields} and their position 
 in the dust sky map shown in figure \ref{virmosindust}.
 
   \begin{table*}
      \caption[]{Survey fields}
         \label{fields}
      \[
        \begin{array}{lllcccc}
           \hline
            \noalign{\smallskip}
            Field      &  \alpha_{2000} & \delta_{2000} &  b & l & E(B-V) & field size\\
                          &  center & center & & & & (I-band)\\
            \noalign{\smallskip}
            \hline
            \noalign{\smallskip}
            0226-04 ``wide'' ({\rm \& XMM-LSS}) &  02{\rm h}24{\rm m}39.75{\rm s} & -04\deg30\arcmin00\arcsec & -58.0 & 172.0 & 0.027 & 5.4\\
           0226-04 ``deep'' & 02{\rm h}26{\rm m}00{\rm s} &-04\deg30\arcmin00\arcsec & & & & 1.45 \\
           1003+01 &  10{\rm h}03{\rm m}39..00{\rm s} & +01\deg54\arcmin39\arcsec & 42.6 & 237.8 & 0.023 & 4.14 \\
           1400+05 &  14{\rm h}00{\rm m}00.00{\rm s} & +05\deg00\arcmin00\arcsec & 62.5 & 342.4 & 0.026 & 4.32 \\
            2217+00 &  22{\rm h}17{\rm m}50.40{\rm s} & +00\deg24\arcmin27\arcsec & -44.0 & 63.3 & 0.062 & 3.60 \\
            \noalign{\smallskip}
            \hline
         \end{array}
      \]
   \end{table*}

\subsection{Survey depth}

The survey has been designed to reach a limiting magnitude $I_{AB}=24.5$ 
(at $5\sigma$ in a 3 arcsec diameter aperture) in
all of the area surveyed, with a smaller 1.3 $deg^2$ area in the 0226-04
field observed to $I_{AB}=25$. The depth of this imaging survey is then at least 
one magnitude deeper than the limiting magnitude(s) of the VIRMOS-VLT 
Deep Survey (VVDS), set to be $I_{AB}=22.5$ for the wide areas and 
$I_{AB}=24$ for the deep area (Le F\`evre et al., 2003), and ensure that
the imaging survey will not introduce any bias in galaxy samples
selected for the spectroscopic survey. 

The actual measured completeness limits are presented in section \ref{depth}


%
%

\subsection{The CFH12K camera}

\begin{figure}
\resizebox{\hsize}{!}{\includegraphics{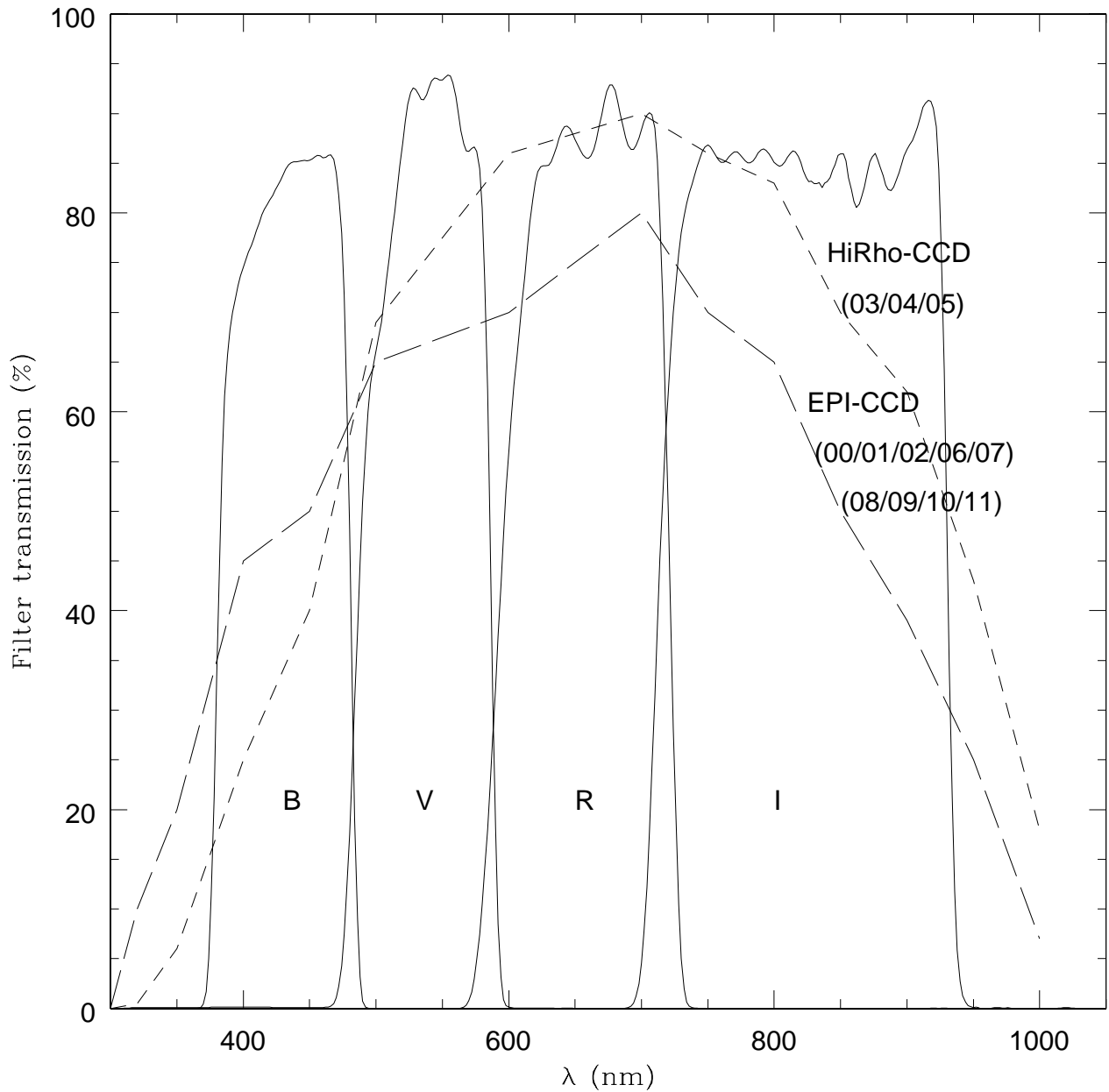}}
   \caption[]{Filter transmission and 
     CCD quantum efficiency of the CFH12K camera used for the 
   VIRMOS optical survey. The camera is composed of 9 epitaxial 
   silicon (EPI) and three high restivity (HiRho) devices. Both devices 
     have different quantum efficiency, as show on the long-dashed and 
      short-dashed curves. The numbers  indicate the CFHT reference number 
        of the CFH12K CCDs.}
\hspace{2cm}
        \label{filter}
\end{figure}

Observations have been carried out with the CFHT-12K camera build by
CFHT and the University of Hawaii \citep{cuillandre00}.
The CFH12K camera   
 ({\tt http://www.cfht.hawaii.edu/Instruments/Imaging/CFH12K/})
is a particularly powerful system. 
 It is based on
a mosaic of 12 3-edge buttable MIT/Lincoln-Labs. CCID20 thinned backside illuminated CCDs, 
  each $2048\times4096$ pixels for a total array size 
  of $12288\times8192$ pixels (18.4 cm $\times$ 12.3 cm in physical size).
The pixel size (15 microns) corresponds to 
  a scale of 0.205 arcsec/pixel at CFHT prime focus, well adapted 
   to the mean seeing at the CFHT prime focus ($\approx 0.8"$).  It covers a field of
view $42\times28$ arcmin$^2$. The camera is composed of two different 
 types of CCDs. Three are high resisitivity chips and nine are 
   epitaxial silicon devices.  The high resistivity CCDs have better 
    efficiency than  the epitaxial beyond 500 nm, but are 
     less efficient in the blue (see figure \ref{filter}). This difference 
       is corrected during the pre-processing step that rescales each CCD
        with respect to a reference device of the camera.

None of the CCDs has a readout noise higher than 6 electrons
 $rms$ nor shows dark current signal over 1 hour time scale, which is largely 
  sufficient for deep imaging surveys with sky background 
    noise dominated images. The readout time 
   is about one minute, which permits to easily split long exposures 
    into short ones  
     without increasing significantly the overheads. In addition to 
     facilitate the construction of master flat fields and fringe 
     patterns and to simplify the cosmic ray rejection process,  
      the image splitting also permits a better control of the camera focus and 
       to reject bad seeing exposures. The autonomy of the 
         LN2 reservoir is much longer than a full single night observation 
	  process and we never detected any temperature drift during the 
	   45 nights dedicated to the VIRMOS survey.

BVRI broad band filters are available on CFH12K. 
 VRI are rather standard Mould filters, but B is special
  (see figure \ref{filter}). They have been 
   designed 
  in order to cover the whole optical spectral range in a continuous manner. 
  The photometric informations have therefore no gaps from one band to 
  the other allowing to probe the whole redshift range from $z=0$ to 
   $z =5$ continuously. This advantage is however obtained at the expense 
     of photometric redshift accuracy for high-$z$ galaxies.

\subsection{Observing strategy}

Each $2\times2$ deg$^2$ field has been divided in a series of CFH12K 
pointings covering the full area, with a slight overlap of $\sim30$ 
arcsec between each pointing. 
A classical shift-and-add observing strategy
has been adopted, each pointing being observed between 
5 and 12 times depending on the filters and depth. 
Because during our first 2 observing 
runs 2 of the 12 CCDs were of poor quality, the
grid of pointings has had to be adjusted to the available pixels for 
these runs. We started operating
with $35\times28$ arcmin$^2$ fields, switching to the 
full $42\times28$ arcmin$^2$
field when new CCDs were installed in the first observing
period of 2000. The field geometry  
is therefore non rectangular in some of the bandpasses.

5by several arcseconds (typically $10-20$ arcsec) between each exposure. 

We aimed to bring first the I-band coverage to $2\times2$deg$^2$
in each of the fields
to ensure that reference catalogs would be available for the magnitude selected
VIRMOS-VLT Deep redshift Survey. We have then completed the coverage
in other bands in a uniform way, proportional to the actual number of
clear nights. 

During the acquisition of a sequence of 
exposures, the aperture flux of control stars was monitored for changes in
photometric conditions. The exposure times were extended in case a loss of
flux was detected during sub-exposures.

\section{Observations}

\subsection{Observed pointings}

The individual pointings observed are listed in 
{\tt http://www.oamp.fr/virmos}, 
with the date of observations, the FWHM
measured on point sources images on the detectors,
and the corresponding CFHT archive files. 

The final CFH12K mosaics resulting from the combination
of these individual pointings are listed in
Table.\ref{TableObsTot} for each of the 4 fields and 4 bands,
together with the resulting exposure times and limiting
magnitudes ($5\sigma$ in a 3 arcsec circular aperture).

\begin{table}
      \caption[]{Total area covered for each of the four survey fields}
         \label{TableObsTot}
\begin{tabular}{lccc}
\hline
Field & Filter & Total Area  & Exp. time   \\
      &        & ($deg^2$)   & (mn) \\  \hline
0230-04 &  B & 3.9 & 120 \\
``Wide'' &  V & 4.17 & 90 \\ 
        &  R & 3.6 & 60\\
        &  I & 5.4 & 60 \\ \hline
0230-04 &  B & 1.45 & 420 \\
``Deep'' &  V & 1.45 & 270 \\ 
        &  R & 1.45 & 180 \\ 
        &  I & 1.45 & 120 \\\hline
1003+01 &  B & 3.0 & 120 \\
        &  V & 2.94 & 90 \\
        &  R & 2.94 & 60 \\
        &  I & 4.14 & 60 \\ \hline
1400+05 &  B & 2.85 & 120 \\
        &  V & 2.85 & 90 \\
        &  R & 3.6 & 60 \\
        &  I & 4.32 & 60 \\ \hline
2217+00 &  B & 1.14 & 120 \\ 
        &  V & 1.14 & 90 \\
        &  R & 1.14 & 60 \\
        &  I & 3.6 & 60 \\ \hline
\hline
\end{tabular}
\end{table}

\subsection{calibrations}

Standard photometric fields from Landolt (1992) have been acquired 
each night survey observations have been obtained.  
  The fields  SA92, SA101 and SA110,  
  have been observed several times over the period 1999-2002 in 
    order to check the reliability of the photometric calibration 
      and the stability of the CFH12K performances. 
The detailed process of photometric calibration, relative 
from exposure to exposure within a set of exposures of a pointing 
in each bandpass, as well as the absolute flux calibration
from the observed standards is described in details in
 \citep{mccracken03}. 
 
\subsection{Image quality}

\begin{figure}
\resizebox{\hsize}{!}{\includegraphics{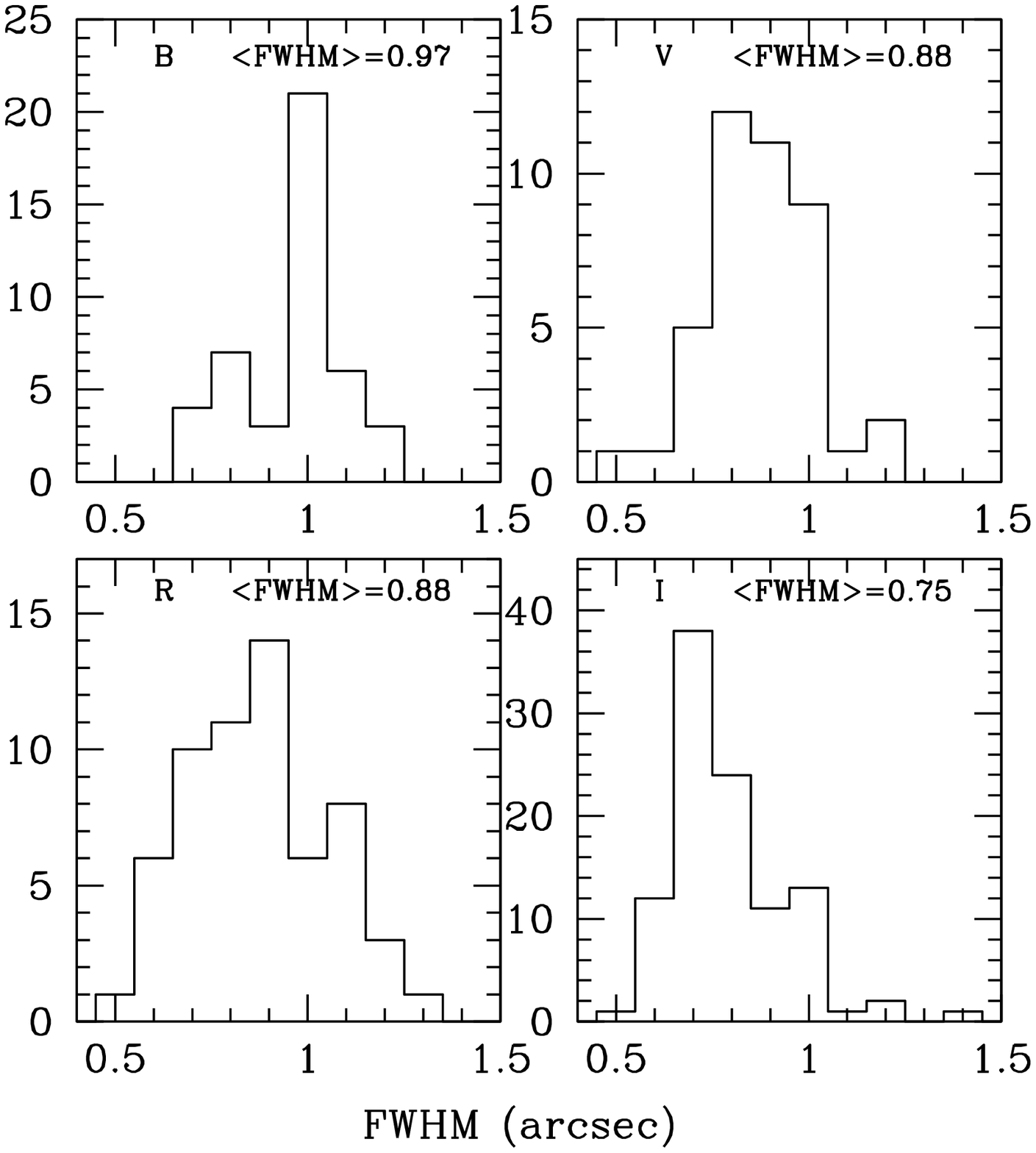}}
   \caption[]{Image quality FWHM measured on point sources for all survey images 
acquired in B (top left), V (top right), R (bottom left), I (bottom right)}
\hspace{5cm}
        \label{FigFWHM}
\end{figure}

The image quality of each exposure acquired for the survey is presented in  
Figure 
\ref{FigFWHM} for each of the bands. The median seeing is FWHM=0.75 arcsec
in the I band but  increases to 0.88 arcsec in V and R, and 0.97 arcsec in B.
  Moreover, the V and R seeing distributions are significantly broader than
    the I and B data sets.
  This trend reflects that atmospheric dispersion is negligible  in 
   I-band and increases towards the B-band.  However, it mostly 
     results from our observing strategy since we preferentially did 
       I-band observations during good seeing periods in order to 
      get high quality I-band selected catalogues for the VMOS spectroscopic 
      sample as well as for the cosmic shear studies.

\section{Data Processing}

\subsection{Pre-processing}

All data are pre-processed using the FLIPS software package (Cuillandre 
et al., 2002).  FLIPS ({\tt http://www.cfht.hawaii.edu/jcc/Flips/flips.html})
  is  composed of C-language programs and Cshell commands that 
   automatically generate  master bias, dark and flat images, build 
    a model of the fringe pattern and subtract
     the master bias,  dark files and fringe pattern as well as 
       the overscan to raw CFH12K images. Each pixel and 
         each image is then rescaled to account for intrinsic efficiency. A 
	  binary mask image provided by CFHT at each observing run 
	  identifies all bad and hot pixels over each CCD of the
	  camera. They are taken into account during the master file
	    generation process. The final data products 
        are pre-calibrated images that includes the description of the 
	 pre-processing history in the FITS header and the 
	  value of the CCD to CCD rescaling coefficient, according to the 
	    different gains of each CCD output and 
	      the quantum efficiency difference between 
	       epitaxial and high resisitivty CCDs. At this stage, all 
	      CCDs are therefore in the same ADU scaling units and
	       arbitrarily normalised to the CCD\#04 efficiency.

	       The FLIPS package has been installed at the Terapix center
	        described below in order to handle the pre-processing 
		 locally.

\subsection{The Terapix facility}

The Terapix facility ({\tt http://terapix.iap.fr}) 
  has been set up as a French national center 
to handle the processing of large imaging cameras. The facility 
includes the hardware and software environment
required to process data on a Terabyte scale. Three COMPACQ 
  XP1000 workstations with up to  2 GB of RAM memory 
 and connected to 4 Terabytes of hard discs 
installed in a secured raid-5 configuration have been
  devoted to the VIRMOS image processing \citep{mellier02}. Over the past four years, Terapix 
  has processed more than 10,000 CFH12K images for the VIRMOS survey, totalling 
     about 2.5 TB of input data. 
A database environment allows to streamline 
data access and storage.

\subsection{Pipeline processing}

Pipeline processing allows efficient data processing with 
excellent quality control. The stable behavior
of the CFH-12K camera allows for very accurate bias, flat field, and 
fringing correction.

The pipeline software has been developped within the Terapix environment.
The main steps are described in \citep{mccracken03} as well as 
  in the Terapix progress reports \citep{mellier02}. The software tools 
    developped for the object detection, image astrometry, photometry, 
      pixel weighting and flaging, image resampling and stacking, 
      object classification and 
        catalogue construction 
	 are described at {\tt http://terapix.iap.fr/soft/} 
        and can be downloaded freely from this site. The 
	  {\tt http://terapix.iap.fr/soft/releases.html} also provides 
	    a documentation for all software tools. 

The astrometric projection of all images into a common system
used to build the large mosaic images has been developed to ensure
an overall absolute calibration to the USNO reference catalogue
 \citep{monet98} accurate to
better than 0.3 arcsecond, while the relative position
accuracy within the catalogs is better than 0.1 arcsec r.m.s.
The photometric calibration ensures that all individual images
are calibrated on the same reference. Photometric zero point uncertainties
are better than 0.1 magnitudes across the final mosaics. A detailled description and 
  application of the pipeline, with a comprehensive 
     description of quality assessments are described in \citep{mccracken03}. 
       Similar procedures applied to the wide VIRMOS 
        survey and the deep U-band survey have been used by Gwyn et al 
	 (2003, in preparation) and Radovich et al (2003), respectively.






\section{Results: survey images and catalogs}

\subsection{Field coverage}

\begin{figure}
\resizebox{\hsize}{!}{\includegraphics{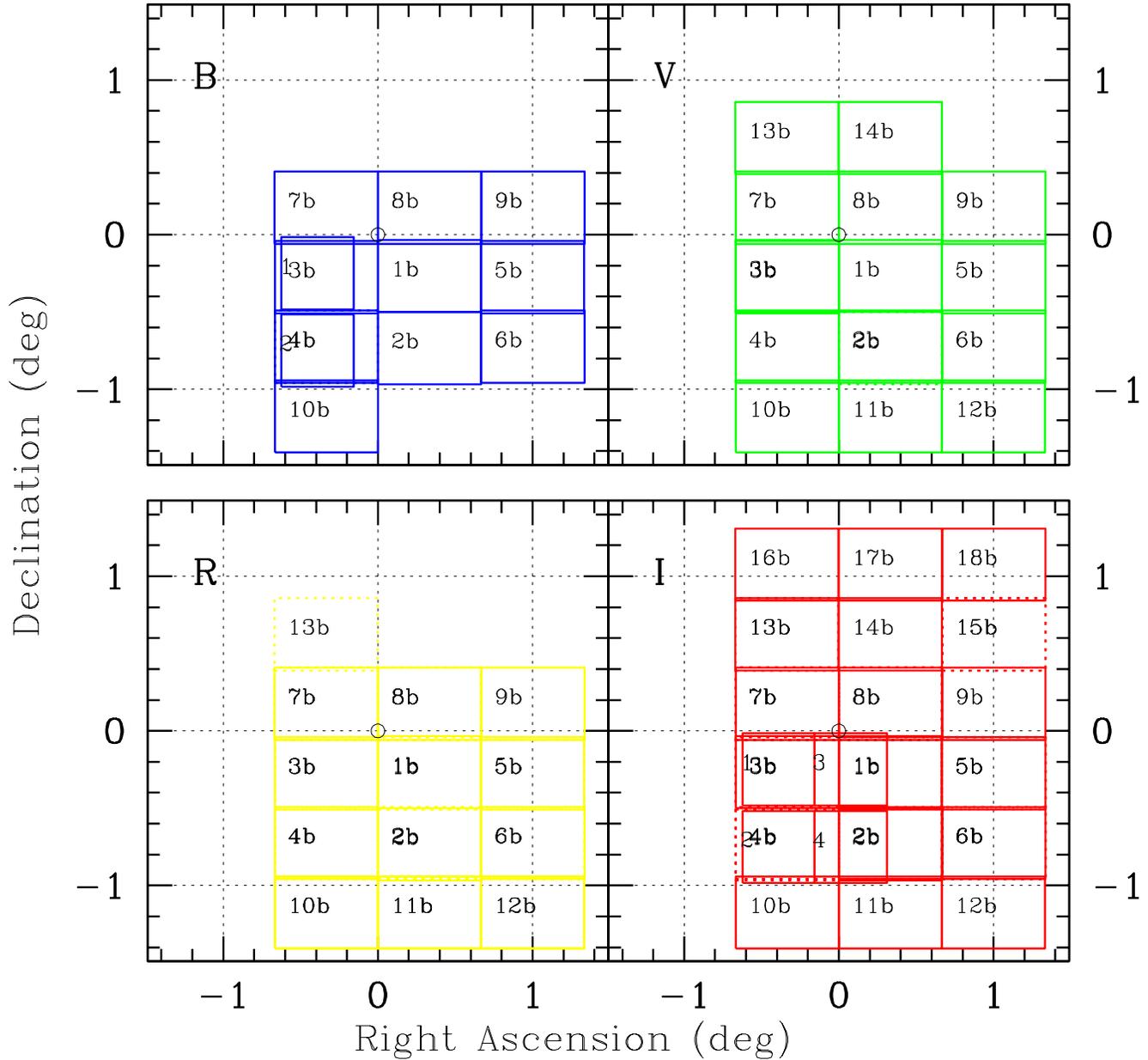}}
   \caption[]{Multi-band coverage of the 0230-04  survey field}
        \label{f02}
\end{figure}

\begin{figure}
\resizebox{\hsize}{!}{\includegraphics{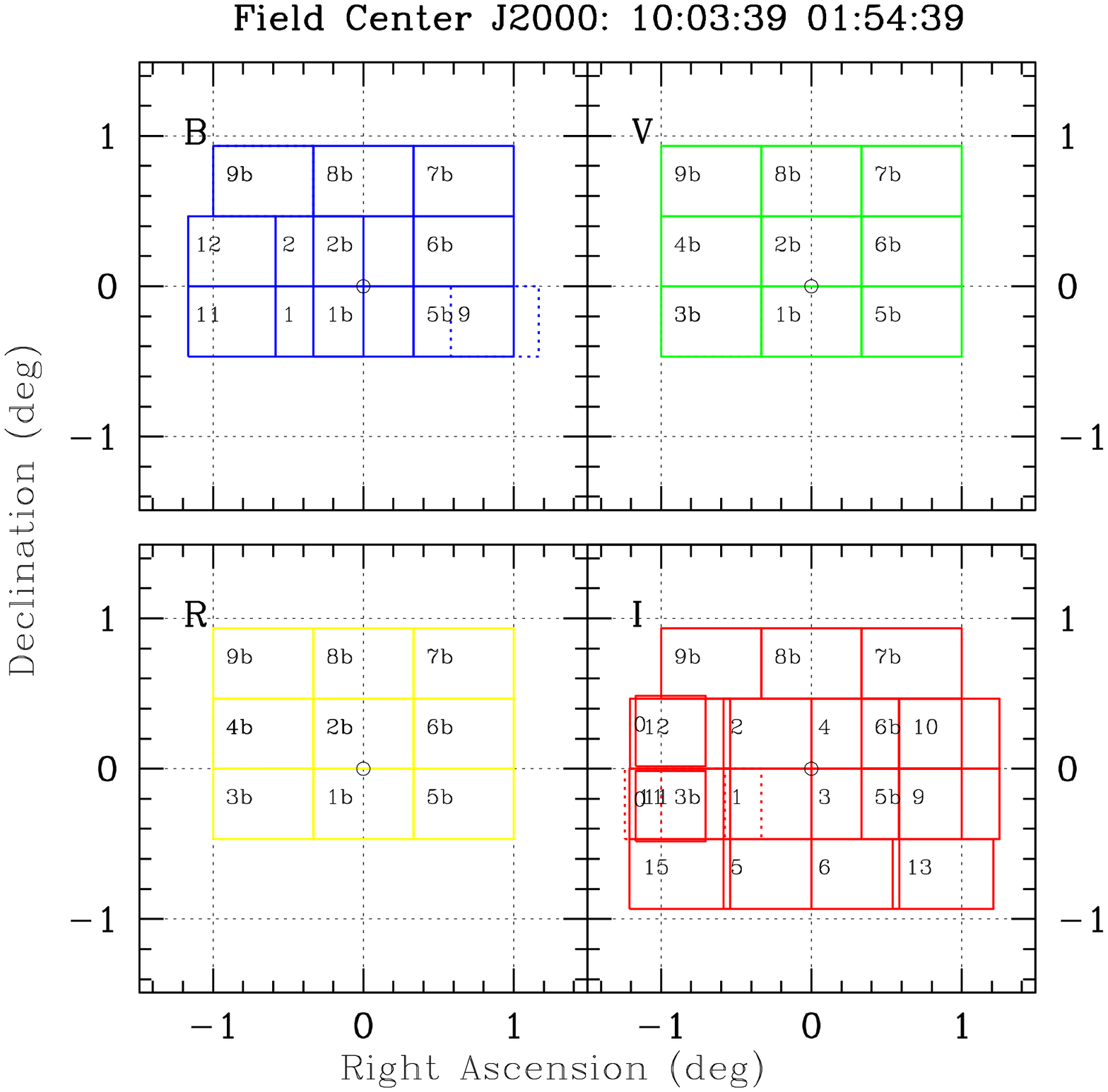}}
   \caption[]{Multi-band coverage of the 1003+01 survey field}
        \label{f10}
\end{figure}

\begin{figure}
\resizebox{\hsize}{!}{\includegraphics{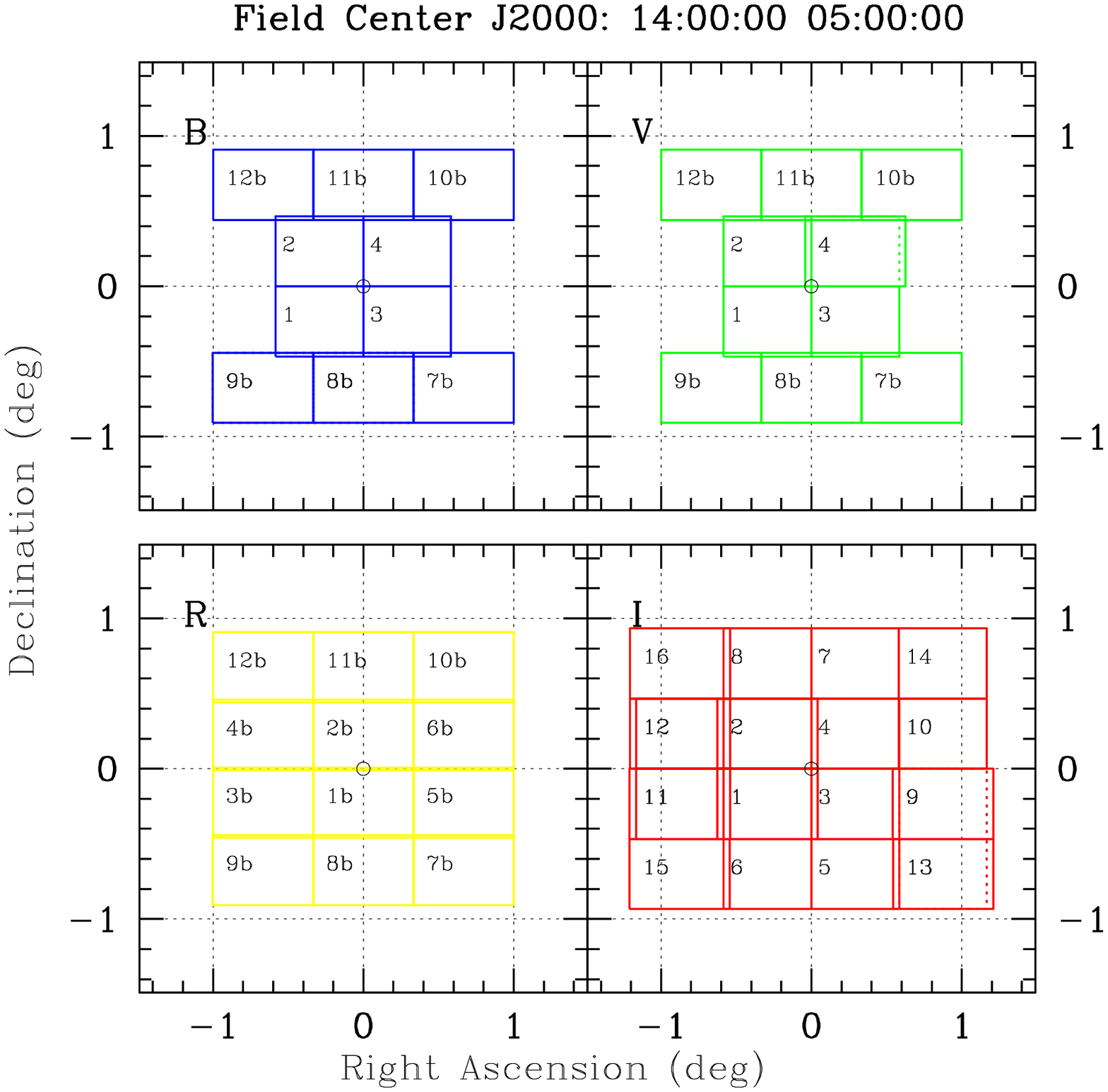}}
   \caption[]{Multi-band coverage of the 1400+05 survey field}
        \label{f14}
\end{figure}

\begin{figure}
\resizebox{\hsize}{!}{\includegraphics{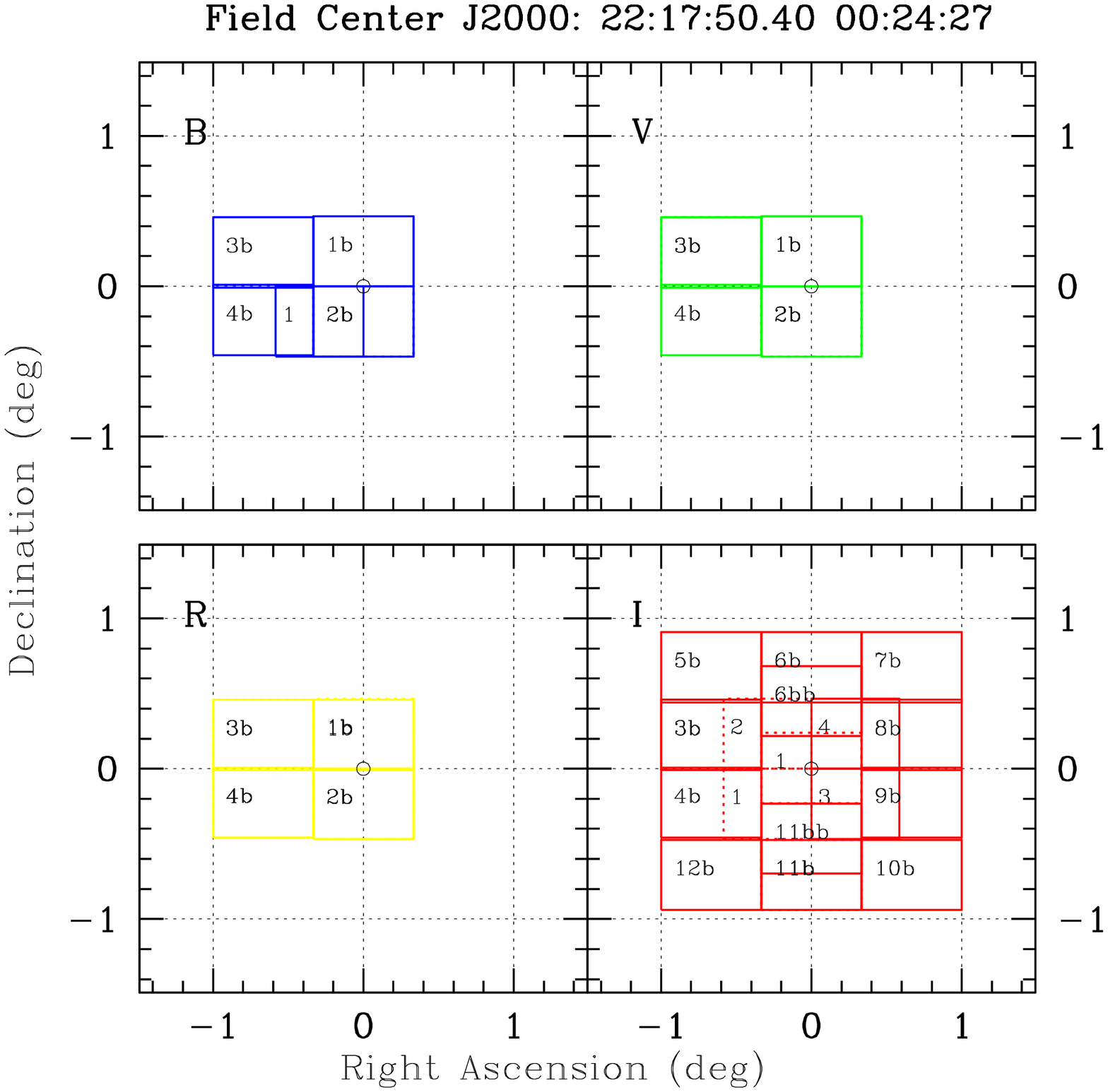}}
   \caption[]{Multi-band coverage of the 2217+00 survey field}
        \label{f22}
\end{figure}

The final multi-wavelength coverage of the four survey fields is shown
in Figure \ref{f02} to Figure \ref{f22}.

\subsection{Catalog content}
\label{depth}

\begin{figure}
\resizebox{\hsize}{!}{\includegraphics{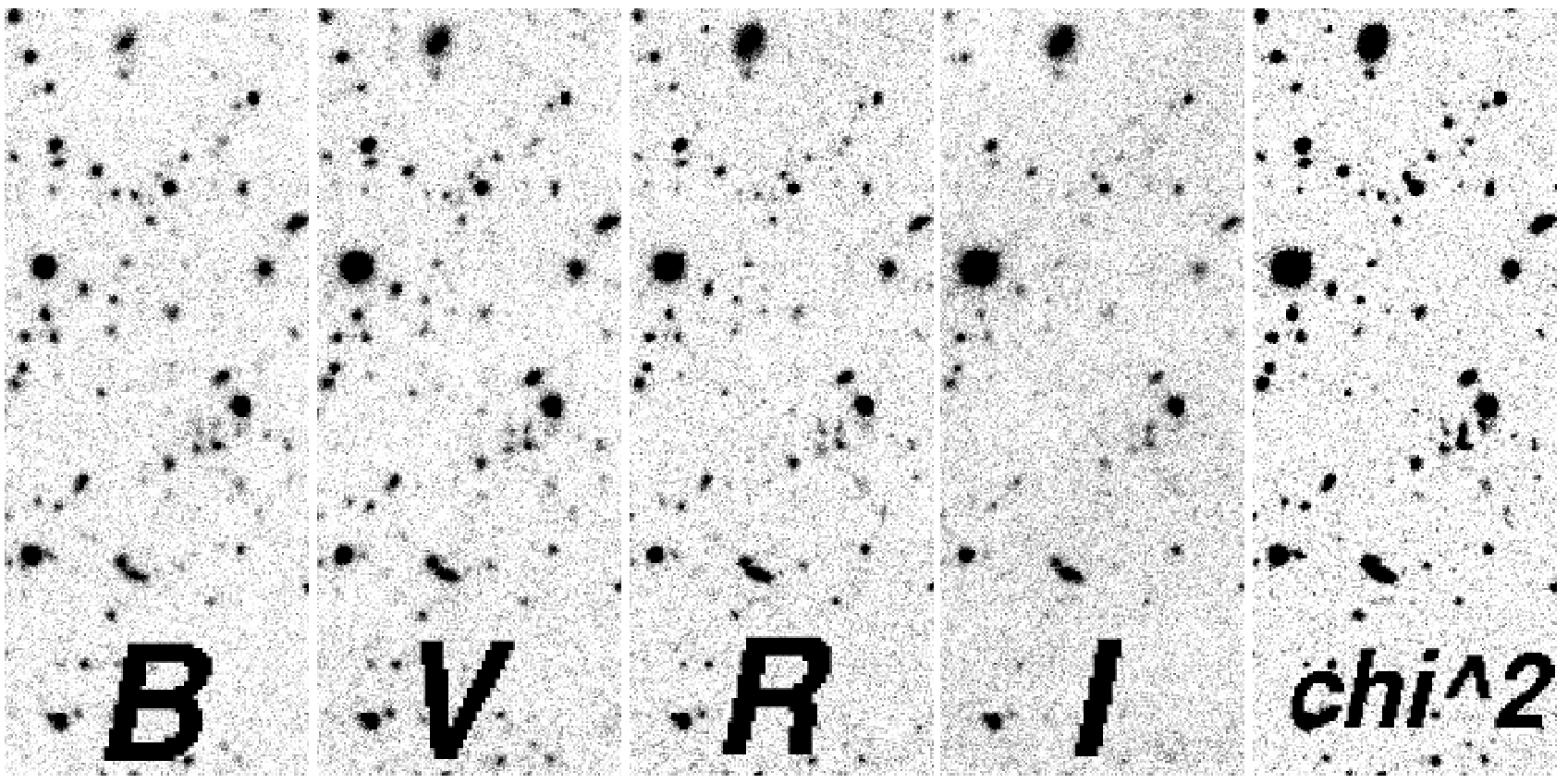}}
   \caption[]{Illustration of the final BVRI and $\chi^2$-images generated 
     for the VIRMOS survey. The $\chi^2$-image is deeper and show sharper 
      detections than the individual filter images.
        }
        \label{chi2image}
\end{figure}

The catalogs are built using the {\sc SExtractor} \citep{bertinarnou96}. Particular care has been taken to ensure uniform source
detection, using a matched xi-squared image (see 
 Fig. \ref{chi2image}) produced from the 
individual band images, or a local threshold algorithm. The 
difference between these two methods has been demonstrated to
be marginal, with no effect in the magnitude range to the
completness limit \citep{mccracken03}.  

Because of the large area covered, bright stars are present in the
mosaics, creating increased background around them. In addition, there
are areas in the images not useable for accurate photometry. These 
can be areas for which the signal to noise is too low
like e.g. areas of overlap between individual CCDs, areas affected
by internal reflexions or ghosts from bright stars around the field,
or even areas  affected by transient effects like satellite trails.
Masks are therefore computed for each
mosaic.

The catalogs contain the following information for each object extracted from 
 the {\sc SExtractor V.2.3} used with the generic configuration files given in 
  the Appendix and which were used as templates for all optical data from the 
    CFH12K part of the survey:
\begin{itemize}
 \item Identifier
 \item Positions: 
        \begin{itemize}
        \item {\tt X-IMAGE} (pixel) , {\tt Y-IMAGE} (pixel), and 
        \item {\tt ALPHA\_J2000} (decimal degrees) and {\tt DELTA\_J2000 } (decimal degrees) 
        \end{itemize}
 \item Flux and Magnitudes in the AB photometric system, defined as 
   -2.5{\rm log}(flux in ADU)+ZP (mag. zero point), including the associated errors: 
       \begin{itemize} 
       \item the total flux and 
             magnitude {\tt FLUX\_AUTO} (ADU), {\tt MAG\_AUTO}, 
       \item the isophotal flux and magnitude,  
             {\tt FLUX\_ISO} (ADU), {\tt MAG\_ISO},  
       \item the  corrected isophotal flux and magnitude
             {\tt FLUX\_ISOCOR} (ADU), {\tt MAG\_ISOCOR},   
       \item the best of {\sc mag\_auto} and {\sc mag\_isocor}, 
             {\tt FLUX\_BEST} (ADU), {\tt MAG\_BEST}, 
       \item the aperture flux and magnitude 
 {\tt FLUX\_APER(2)} (ADU), {\tt MAG\_APER(2)}, 
       \end{itemize}
Colors can then be derived directly using the aperture magnitudes
computed within the same aperture for all bands.
 \item Shapes: 
       \begin{itemize} 
       \item image size, 
  {\tt KRON\_RADIUS} (pixel) and {\tt ISOAREA\_IMAGE} (pixel$^2$), 
       \item major axis {\tt A\_IMAGE} (pixel) , {\tt A\_WORLD} (decimal degrees), 
       \item minor axis {\tt B\_IMAGE} (pixel), {\tt B\_WORLD} (decimal degrees) , 
       \item position angle of main  axes (all angles are measured counter-clockwise from the east. 
  {\sc theta} denotes the position angle between the major axis, "{\sc a}",
    and the {\sc naxis1} (east-west) image axis),
       {\tt THETA\_IMAGE} (decimal degrees), {\tt THETA\_J2000 } (decimal degrees),
       \item the values of the values of the weighted second moment matrix coefficients, 
             {\tt X2\_IMAGE} (pixel$^2$), {\tt Y2\_IMAGE} (pixel$^2$), 
	     {\tt XY\_IMAGE} (pixel$^2$), 
        \item {\tt X2\_WORLD} (decimal degees$^2$), 
              {\tt Y2\_WORLD} (decimal degees$^2$), {\tt XY\_WORLD}
              (decimal degees$^2$), 
        \item peak surface brightness {\tt MU\_MAX} (mag/arcsec$^2$), 
	\item detection threshold above the background {\tt MU\_THRESHOLD} (mag/arcsec$^2$). 
	\end{itemize}
 \item Flags: several flags ({\tt FLAGS})  are stored for each object.  
  they indicate:
        \begin{itemize}
	\item saturated pixels,
	\item pixel located inside a masked area, as defined above.
        \item Star-Galaxy classification: 
   A star / galaxy classifier defined from the half-light radius
vs. magnitude diagram (see \citep{mccracken03}). Stars are selected along 
  the vertical branch at fixed radius corresponding to seeing disk size.
   This method initally described in \citep{falhman94} turns out to be 
    more reliable than the initial {\tt CLASS\_STAR} of {\sc SExtractor},
        \end{itemize}
\end{itemize}

The photometry and astrometry quality have been assessed through many 
quality checks.
 The positional accuracy has been checked
against the USNO catalog. The photometry accuracy has been
first assessed through
the Elixir queue observing program part of the queue observations
at CFHT, which monitors the photometric zero points accross nights.
The magnitudes of objects common to adjacent CCDs have been compared
and showed differences in the range 0.05--0.1 magnitudes. 
The comparison of colors for the bright stellar objects in the fields
with the observed and predicted locus of stars has shown 
color errors less than 0.1 magnitudes. Counts of galaxies 
are in full agreement with the litterature. Finally, the angular 
correlation function shows similar shape and amplitude
as from other imaging surveys. All these quality checks
are extensively described in \citep{mccracken03}. 
 They ensure that the data are free of 
systematics before science analysis can be conducted.
 
In the I band, a total of 2.175 million objects have been
detected. Among these, 1.15 million objects have $17.5\leq I_{AB} \leq 24$,
in the range of the VIRMOS-VLT Deep Survey. 

\begin{figure*}
\begin{center}
\resizebox{\hsize}{!}{\includegraphics{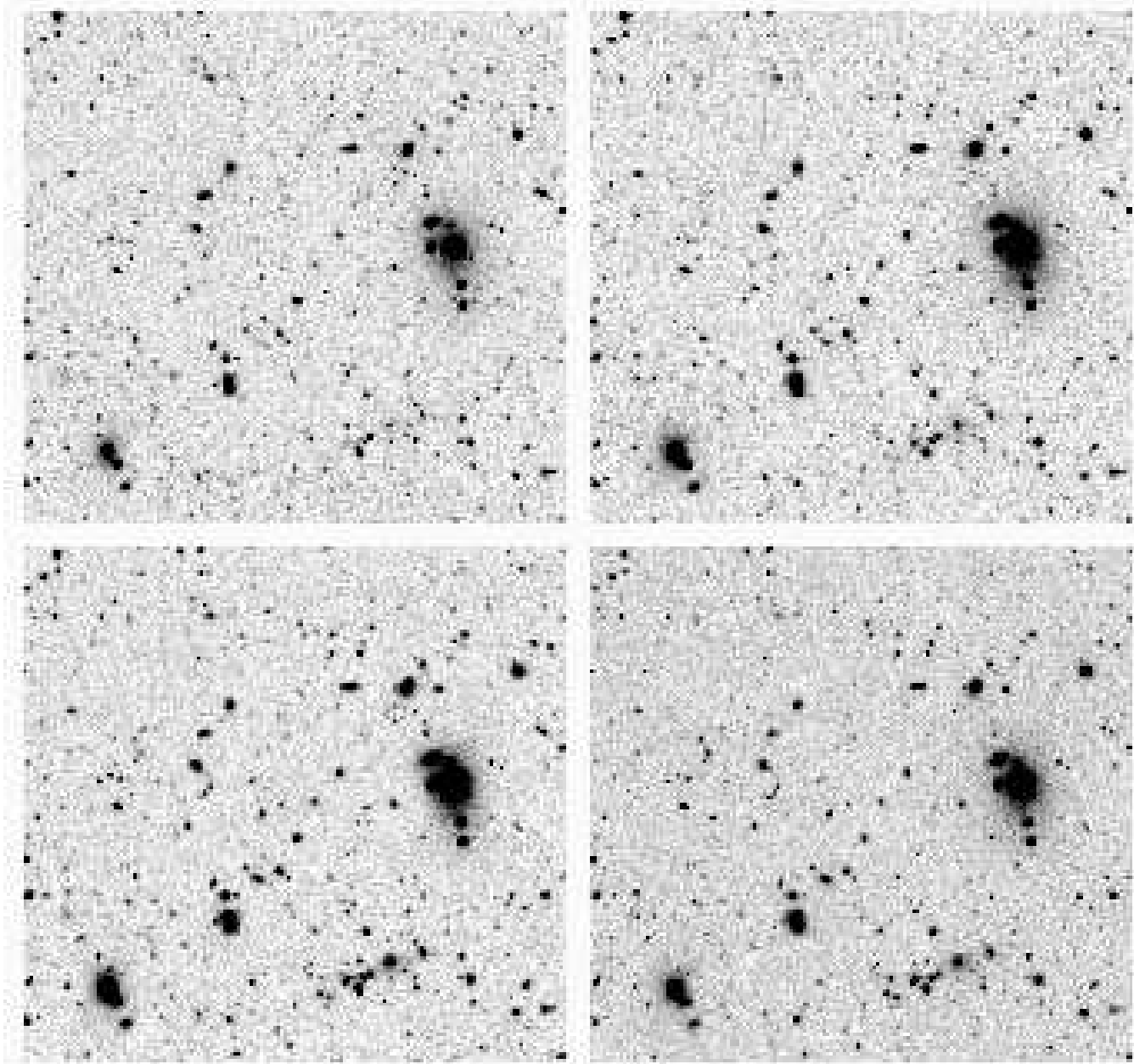}}
  \caption[]{From upper left, clockwise: 
             B,V,R, and I images of a $3.4\times3.4$arcmin$^2$
             area in the 0226-04 ``Deep field''}
        \label{imagesBVRI}
\end{center}
\end{figure*}

An example of a multi-color set is given in Figure \ref{imagesBVRI}.


\subsection{Database access}

All catalogs and mosaic images are stored in an interactive database 
implemented under the Oracle-8 environment. Specific
database development have been conducted to allow
for easy database query and data retrieval. 

The catalogs are now in direct access to the CFHT-VIRMOS
consortium. They will be released for general use 
for CFHT users starting July 1st, 2003 ({\tt http://www.oamp.fr/virmos}).
To access the database, first a password should be requested
sending an email to {\tt virmos.database@oamp.fr}. 

The database
allows to query the photometric catalog and to retrieve 
image sections. Any parameter listed in the photometric catalog
can be used to define a user selection, e.g. magnitudes, colors,
and/or positions. A user catalog is then produced with entries
as selected by the user from the list of catalog parameters.

\section{Conclusions}

A deep imaging survey has been conducted in 4 high
galacitic latitude areas with the CFH-12K camera. 
A total area of more than $17$deg$^2$ has been imaged
in I band, and significant coverage of
the survey area has been performed also in B, V and R. 

 Pipeline processing has been conducted at the Terapix
facility at IAP to produce large mosaic images
calibrated in astrometry and photometry.

Photometric catalogs containing positions, magnitudes,
colors, shapes of more than 2 million objects
have been produced for more than 17deg$^2$ 
and quality control has been applied
as described in joint papers. These catalogs and
mosaic images will
be released for general use starting July 1st, 2003,
at {\tt http://www.oamp.fr/virmos}.

While many scientific programs are under way, the survey  
 has already been used for cosmic shear. The {\sc virmos-descart cfht} survey 
 ({\tt http://www.terapix.iap.fr/Descart/cfhtsurvey.html}) primarily focusses 
  on the I-band 
   sample to measure galaxy shape with high accuracy and probe the dark 
    matter properties and cosmological parameters.  Because these cosmic 
    shear results provide also an estimate of systematics on galaxy shapes
      and can be cross-checked with independent cosmic shear results 
        from other teams, they also provide quality assessments on the
	  VIRMOS data set. In addition to 
    the detection of 
      cosmic shear signal
      \citep{vanwaerbeke00}, constraints on $\Omega_m$, 
    $\sigma_8$ \citep{vanwaerbeke02}, the biasing properties of the 
     dark matter \citep{hoekstra02,pen03a} as 
     well is its non-Gaussian properties 
      \citep{bernardeau02, pen03b} and the 3-dimension 
     power spectrum of the dark matter 
      \citep{pen03a} have already been provided 
      using the VIRMOS survey for cosmic shear. 
      
The VIMOS spectroscopic survey is now started at the ESO/VLT 
 and more than 20,000 spectra have been collected
  \citep{lefevre03}. The joint 
   U-band, optical, near infrared and spectroscopic informations
     are now under process to probe the star formation history,
       the clustering history of galaxies up to redshift $z \approx $ 5 
        and the properties of the light and dark matter relations.

\begin{acknowledgements}
We thank the CFHT time allocation committee for continuous 
support of this long term program, the CFHT staff for 
the execution of the observations performed in queue
scheduling mode, the Terapix staff for its 
  continuous help during the VIRMOS image 
    processing period, the CNRS-INSU, CEA/DAPNIA and 
 the Programme National de Cosmologie for support of the Terapix
facility and IAP and OAMP for funding of this program.
\end{acknowledgements}

\newpage
\section*{A1: {\sc SExtractor} configuration files}
Tables \ref{sexchi2} and \ref{sexwide} list the parameters 
  used for all
 $\chi^2$-images and all wide field analysis of VIRMOS optical 
 obervations with the CFH12K.

   \begin{table*}
      \caption[]{{\sc SExtractor} configuration fiels used for the 
F02 deep survey, $\chi^2$-image}
         \label{sexchi2}
      \[
\begin{tabular}{lll}
           \hline
            \noalign{\smallskip}
\multicolumn{3}{l}{{\tt \#Default configuration file for} {\sc SExtractor V1.2b14 }}  \\
{\tt \# EB 23/07/98} & & \\
\multicolumn{3}{l}{\tt \# (*) indicates parameters which can be omitted from this config file.} \\
\multicolumn{3}{l}{\tt \#----------------------------------------- Catalog ------------------------------------}
 \\

CATALOG\_NAME  &  02hr\_chisq.cat &\# name of the output catalog \\
CATALOG\_TYPE  &  ASCII\_HEAD  &       \# "NONE","ASCII\_HEAD","ASCII","FITS\_1.0" \\
                        & &       \# or "FITS\_LDAC" \\

PARAMETERS\_NAME& photom\_new.param&  \# name of the file \\
                            & &   \# containing catalog contents \\

\multicolumn{3}{l}{\tt \#---------------------------------------- Extraction ----------------------------------} \\

DETECT\_TYPE    & CCD &            \# "CCD" or "PHOTO" (*) \\
DETECT\_MINAREA & 3   &            \# minimum number of pixels above threshold\\
DETECT\_THRESH  & 1.1 &		\# $<$sigmas$>$ or $<$threshold$>$,$<$ZP$>$ in mag.arcsec$^{-2}$\\
ANALYSIS\_THRESH& 1.1 &		\# $<$sigmas$>$ or $<$threshold$>$,$<$ZP$>$ in mag.arcsec$^{-2}$\\
THRESH\_TYPE    & ABSOLUTE & \\
FILTER         & N & \\
FILTER\_NAME    & gauss\_3.0\_7x7.conv & \# name of the file containing the filter\\

DEBLEND\_NTHRESH& 32  &            \# Number of deblending sub-thresholds\\
DEBLEND\_MINCONT& 0.002 &          \# Minimum contrast parameter for deblending\\

CLEAN          & Y &  \\
CLEAN\_PARAM    & 1.0 &             \# Cleaning efficiency\\
MASK\_TYPE      & CORRECT &        \# type of detection MASKing: can be one of\\
               & &                 \# "NONE", "BLANK" or "CORRECT"\\

\multicolumn{3}{l}{\tt \#--------------------------------------- Photometry -----------------------------------} \\

PHOT\_APERTURES &   15.,25.&  	\# MAG\_APER aperture diameter(s) in pixels\\
PHOT\_FLUXFRAC &  0.2, 0.5, 0.8&  	\# Fraction of FLUX\_AUTO defining FLUX\_RADIUS\\
PHOT\_AUTOPARAMS& 2.5, 3.5     &   \# MAG\_AUTO parameters: $<$Kron\_fact$>$,$<$min\_radius$>$\\
SATUR\_LEVEL    & 40000.        &  \# level (in ADUs) at which arises saturation\\
PHOT\_AUTOAPERS & 15.0, 15.0   &   \# MAG\_AUTO minimum apertures: estimation,photometry\\

MAG\_ZEROPOINT  & 31.90        &   \# magnitude zero-point\\
MAG\_GAMMA      & 4.0          &   \# gamma of emulsion (for photographic scans)\\
GAIN           & 2.0	   &	\# detector gain in e$-$/ADU.\\
PIXEL\_SCALE    & 0.205  & 	\# size of pixel in arcsec (0=use FITS WCS info).\\

\multicolumn{3}{l}{\tt \#--------------------------------- Star/Galaxy Separation ----------------------------} \\

SEEING\_FWHM    & 0.9 &                               \# stellar FWHM in arcsec\\
STARNNW\_NAME   & default.nnw & \# Neural-Network\_Weight table filename\\

\multicolumn{3}{l}{\tt \#------------------------------------- Background ------------------------------------} \\

BACK\_SIZE      & 128         &    \# Background mesh: $<$size$>$ or $<$width$>$,$<$height$>$\\
BACK\_FILTERSIZE& 9           &    \# Background filter: $<$size$>$ or $<$width$>$,$<$height$>$\\
BACKPHOTO\_TYPE & LOCAL       &    \# can be "GLOBAL" or "LOCAL" (*)\\
BACKPHOTO\_THICK& 30          &    \# thickness of the background LOCAL annulus (*)\\
BACK\_TYPE      & AUTO   & \\
BACK\_VALUE     & 0 & \\
\multicolumn{3}{l}{\tt \#------------------------------------- Check Image -----------------------------------} \\

CHECKIMAGE\_TYPE &NONE  & \# can be one of "NONE", "IDENTICAL", \\
                    &   &          \# "BACKGROUND", "-BACKGROUND","BACKGROUND\_RMS",\\

CHECKIMAGE\_NAME &check.fits &      \# Filename for the check-image (*)\\

\multicolumn{3}{l}{\tt \#-------------------------------- Memory (change with caution!) ---------------------} \\
MEMORY\_OBJSTACK& 2000      &      \# number of objects in stack\\
MEMORY\_PIXSTACK& 5000000   &      \# number of pixels in stack\\
MEMORY\_BUFSIZE & 512       &      \# number of lines in buffer\\

\multicolumn{3}{l}{\tt \#----------------------------------- Miscellaneous ---------------------------------} \\

VERBOSE\_TYPE  &  NORMAL      &     \# can be "QUIET", "NORMAL" or "FULL" (*)\\

\multicolumn{3}{l}{\tt \#------------------------------------- New Stuff -----------------------------------} \\

WEIGHT\_TYPE   & MAP\_WEIGHT &  \\
\#WEIGHT\_TYPE  &  BACKGROUND & \\
\noalign{\smallskip}
            \hline
         \end{tabular}
      \]
   \end{table*}
\newpage
   \begin{table*}
      \caption[]{{\sc SExtractor} configuration files used for the 
wide field survey}
         \label{sexwide}
      \[
\begin{tabular}{lll}
           \hline
            \noalign{\smallskip}
\multicolumn{3}{l}{{\tt \#Default configuration file for} {\sc SExtractor V1.2b14 }}  \\
{\tt \# EB 23/07/98} & & \\
\multicolumn{3}{l}{\tt \# (*) indicates parameters which can be omitted from this config file.} \\
\multicolumn{3}{l}{\tt \#----------------------------------------- Catalog ------------------------------------}
 \\

CATALOG\_NAME  &  F22It3.cat &\# name of the output catalog \\
CATALOG\_TYPE  &  ASCII\_HEAD  &       \# "NONE","ASCII\_HEAD","ASCII","FITS\_1.0" \\
                        & &       \# or "FITS\_LDAC" \\

PARAMETERS\_NAME& header.param&  \# name of the file \\
                            & &   \# containing catalog contents \\

\multicolumn{3}{l}{\tt \#---------------------------------------- Extraction ----------------------------------} \\

DETECT\_TYPE    & CCD &            \# "CCD" or "PHOTO" (*) \\
DETECT\_MINAREA & 3   &            \# minimum number of pixels above threshold\\
DETECT\_THRESH  & 1 &		\# $<$sigmas$>$ or $<$threshold$>$,$<$ZP$>$ in mag.arcsec$^{-2}$\\
ANALYSIS\_THRESH& 1 &		\# $<$sigmas$>$ or $<$threshold$>$,$<$ZP$>$ in mag.arcsec$^{-2}$\\
THRESH\_TYPE    & RELATIVE & \\
FILTER         & Y & \\
FILTER\_NAME    & gauss\_3.0\_7x7.conv & \# name of the file containing the filter\\

DEBLEND\_NTHRESH& 32  &            \# Number of deblending sub-thresholds\\
DEBLEND\_MINCONT& 0.002 &          \# Minimum contrast parameter for deblending\\

CLEAN          & Y &  \\
CLEAN\_PARAM    & 1.0 &             \# Cleaning efficiency\\
MASK\_TYPE      & CORRECT &        \# type of detection MASKing: can be one of\\
               & &                 \# "NONE", "BLANK" or "CORRECT"\\

\multicolumn{3}{l}{\tt \#--------------------------------------- Photometry -----------------------------------} \\

PHOT\_APERTURES &   15.,25.&  	\# MAG\_APER aperture diameter(s) in pixels\\
PHOT\_FLUXFRAC &  0.2, 0.5, 0.8&  	\# Fraction of FLUX\_AUTO defining FLUX\_RADIUS\\
PHOT\_AUTOPARAMS& 2.5, 3.5     &   \# MAG\_AUTO parameters: $<$Kron\_fact$>$,$<$min\_radius$>$\\
SATUR\_LEVEL    & 1200.        &  \# level (in ADUs) at which arises saturation\\
PHOT\_AUTOAPERS & 15.0, 15.0   &   \# MAG\_AUTO minimum apertures: estimation,photometry\\

MAG\_ZEROPOINT  & 30.00        &   \# magnitude zero-point\\
MAG\_GAMMA      & 4.0          &   \# gamma of emulsion (for photographic scans)\\
GAIN           & 0.04	   &	\# detector gain in e$^-$/ADU.\\
PIXEL\_SCALE    & 0.  & 	\# size of pixel in arcsec (0=use FITS WCS info).\\

\multicolumn{3}{l}{\tt \#--------------------------------- Star/Galaxy Separation ----------------------------} \\

SEEING\_FWHM    & 0.9 &                               \# stellar FWHM in arcsec\\
STARNNW\_NAME   & default.nnw & \# Neural-Network\_Weight table filename\\

\multicolumn{3}{l}{\tt \#------------------------------------- Background ------------------------------------} \\

BACK\_SIZE      & 512         &    \# Background mesh: $<$size$>$ or $<$width$>$,$<$height$>$\\
BACK\_FILTERSIZE& 9           &    \# Background filter: $<$size$>$4 or $<$width$>$,$<$height$>$\\
BACKPHOTO\_TYPE & LOCAL       &    \# can be "GLOBAL" or "LOCAL" (*)\\
BACKPHOTO\_THICK& 30          &    \# thickness of the background LOCAL annulus (*)\\
BACK\_TYPE      & AUTO   & \\
BACK\_VALUE     & 0 & \\
\multicolumn{3}{l}{\tt \#------------------------------------- Check Image -----------------------------------} \\

CHECKIMAGE\_TYPE &NONE  & \# can be one of "NONE", "IDENTICAL", \\
                    &   &          \# "BACKGROUND", "-BACKGROUND","BACKGROUND\_RMS",\\
                    &    &         \# "MINIBACKGROUND", "MINIBACK\_RMS",\\
                    &     &        \# "FILTERED", "OBJECTS", "-OBJECTS",\\
                    &      &       \# "SEGMENTATION", or "APERTURES",\\

CHECKIMAGE\_NAME &check.fits &      \# Filename for the check-image (*)\\

\multicolumn{3}{l}{\tt \#-------------------------------- Memory (change with caution!) ---------------------} \\
MEMORY\_OBJSTACK& 2000      &      \# number of objects in stack\\
MEMORY\_PIXSTACK& 5000000   &      \# number of pixels in stack\\
MEMORY\_BUFSIZE & 512       &      \# number of lines in buffer\\

\multicolumn{3}{l}{\tt \#----------------------------------- Miscellaneous ---------------------------------} \\

VERBOSE\_TYPE  &  NORMAL      &     \# can be "QUIET", "NORMAL" or "FULL" (*)\\

\multicolumn{3}{l}{\tt \#------------------------------------- New Stuff -----------------------------------} \\

WEIGHT\_TYPE   & MAP\_WEIGHT &  \\
\#WEIGHT\_TYPE  &  BACKGROUND & \\
\noalign{\smallskip}
            \hline
         \end{tabular}
      \]
   \end{table*}
\end{document}